\documentclass[preprint, superscriptaddress, 
 amsmath, amssymb, aps, prb,  
 ]{revtex4-2}

\usepackage{graphicx}
\usepackage{dcolumn}
\usepackage{amsmath, amssymb,bm}
\usepackage[none]{hyphenat}
\usepackage{verbatim}
\usepackage{textcomp}
\usepackage{subcaption}
\usepackage{hyperref}
\hypersetup{colorlinks=true, urlcolor=blue, linkcolor=blue, citecolor=blue}
\begin{document}

\preprint{}

\title{Investigation of Cu-site disorder in undoped and doped BiCuSeO}

\author{Sheetal~Jain}
\author{Christopher~J.~S.~Heath}
\author{Dixshant~Shree~Shreemal}
\author{Blair~W.~Lebert}
\affiliation{University of Toronto, Toronto, Ontario, M5S 1A7, Canada}
\author{Ning~Chen}
\author{Weifeng~Chen}
\affiliation{Canadian Light Source (CLS), Saskatoon, Saskatchewan, S7N 2V3, Canada}
\author{Young-June~Kim}
\email{youngjune.kim@utoronto.ca}
\affiliation{University of Toronto, Toronto, Ontario, M5S 1A7, Canada}

\date{\today}

\begin{abstract}
We carried out X-ray diffraction and Extended X-ray Absorption Fine Structure (EXAFS) studies to investigate the origin of the low lattice thermal conductivity in BiCuSeO, and the role of silver (Ag) doping in doped samples. BiCuSeO is a promising thermoelectric material with high thermoelectric efficiency, which is significantly enhanced by doping either single Ag dopants or dual dopants (Pb and Ag). We verified that the thermal displacement parameters associated with copper (Cu) are significantly large in undoped BiCuSeO. Ag dopant, which replaces Cu, was also found to have similarly large thermal displacement parameters, that remain large down to low temperatures. Our results point towards significant disorder on Cu-site in both undoped and doped BiCuSeO, which is retained by Ag dopant on replacing Cu. The disorder is observed to be localized on the Cu-site and seems to be independent of other atoms in the crystal structure. Our observation of the disorder, which could be either static or quasi-static, is consistent with a ``rattling'' mode scenario. 
\end{abstract}


\maketitle


\section{\label{sec:intro}Introduction}

Thermoelectric effects directly convert thermal energy into electrical energy, and vice-versa. They have potential applications in energy generation from waste heat, and refrigeration, and may play a major role in alleviating the looming energy crisis and global climate change issues~\cite{Tritt1999, Bell2008, Snyder2008, Zhu2017}. Highly effective thermoelectric materials must have high electrical conductivity and low thermal conductivity. Since electronic thermal conductivity is linked to electrical conductivity by the Wiedemann–Franz relation, the main goal of designing thermoelectric materials is to optimize electrical transport while limiting lattice thermal conductivity. 

\begin{figure}[htbp]
\centering
\includegraphics[width=0.5\textwidth]{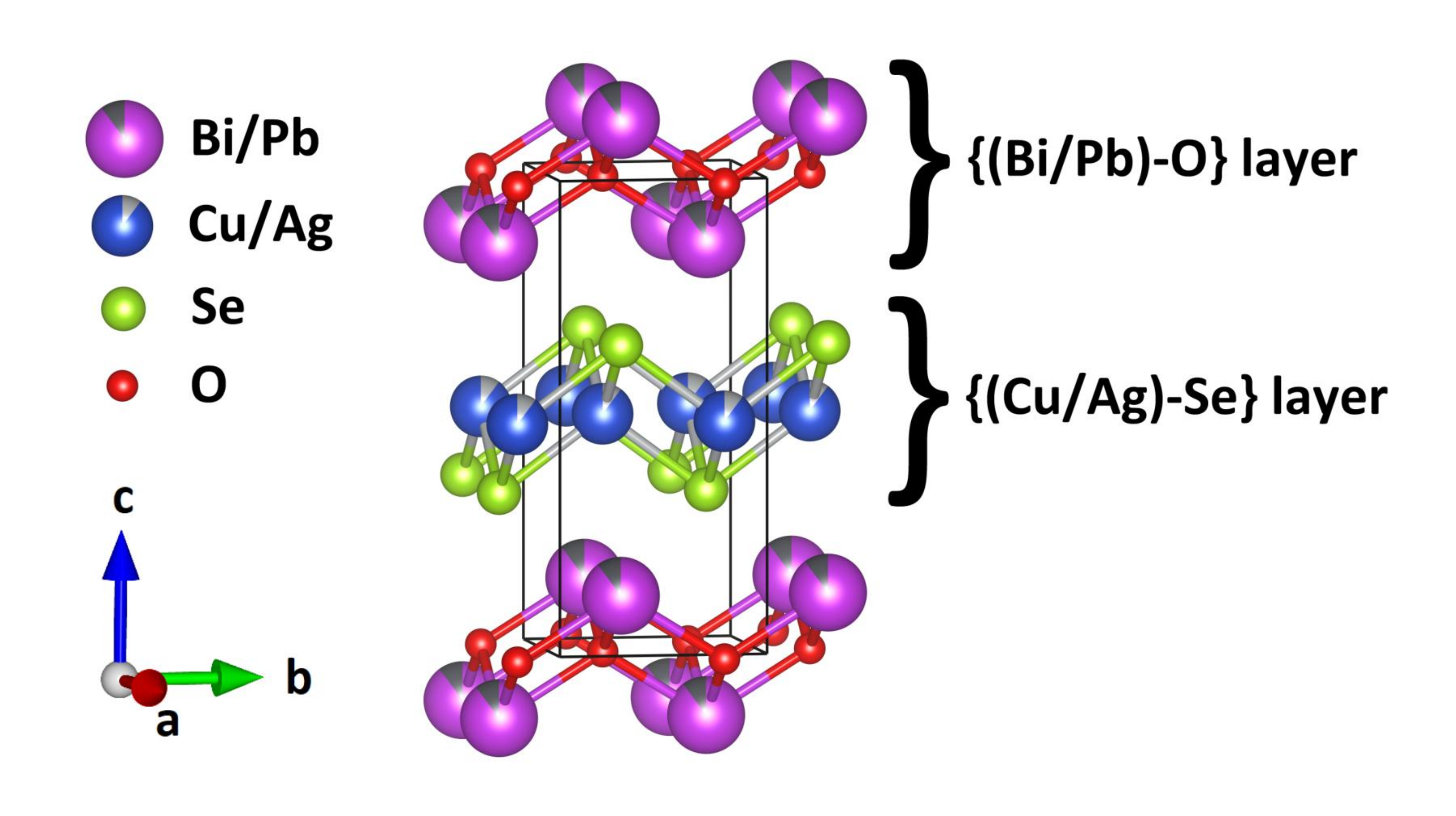}
\caption{\label{fig:fig1_structure}Schematic representation of the layered crystal structure of Bi$_{1-x}$Pb$_{x}$Cu$_{1-x}$Ag$_{x}$SeO, with Pb dopants going to Bi-sites and Ag dopants going to Cu-sites. The tetragonal unit cell, with the space group P4/nmm (No. 129), is shown with the solid black line. Crystal structure visualization was done using the VESTA software~\cite{Momma2011}.}
\end{figure}

One of the promising examples of thermoelectric materials is BiCuSeO. Its crystal structure consists of alternating layers of fluorite-like [Bi$_2$O$_2$]$^{2+}$ oxide layer and anti-fluorite-like [Cu$_2$Se$_2$]$^{2-}$ chalcogenide layer, stacked along the c-axis~\cite{Kusainova1994}, as shown in Figure~\ref{fig:fig1_structure}. This layered structure is similar to the famous iron-based superconductor, LaFeAsO~\cite{Zhao2014, Kamihara2008}. The oxide layer is insulating, with ionic bonds, while the conducting chalcogenide layer consists of covalent bonds~\cite{Zhao2010}. BiCuSeO has a moderate Seebeck coefficient and an unusually low lattice thermal conductivity~\cite{Zhao2010}. While undoped BiCuSeO exhibits low electrical conductivity, it can be significantly enhanced by doping. BiCuSeO has been doped with various alkaline earth metals~\cite{Zhao2010,Barreteau2012,JLi2012,Pei2013,FLi2013,JLi2013,Lan2013}, alkali metals~\cite{SunLee2013,JLi2014,Achour2018}, transition metals~\cite{Liu2015,Farooq2016,Das2019,LiuAg2015}, rare earth metals~\cite{Feng2019,Novitskii2019,Feng2020,Kang2020,BFeng2020} and other post-transition metals~\cite{Pan2013,Luu2013,YcLiu2013,JLLan2013,Yang2017,Das2017,Feng2017,Lei2019,Xu2021}, with dopants replacing bismuth (Bi) to generate holes in p-type BiCuSeO. In particular, lead (Pb) was observed to lead to the highest enhancement of electrical conductivity~\cite{JLLan2013}, while calcium (Ca) doping led to the lowest thermal conductivity~\cite{Pei2013}. The enhanced electrical conductivity, coupled with the moderate Seebeck coefficient and intrinsic low lattice thermal conductivity of BiCuSeO leads to high thermoelectric efficiency in hole-doped BiCuSeO~\cite{FLi2012}. To further optimize thermoelectric performance, substitution of copper (Cu) with transition metals was employed, which was observed to reduce lattice thermal conductivity even further~\cite{Tan2014,Luu2014,Farooq2017,Sun2017,Ren2014}. Furthermore, electrical and thermal transport can be optimized synergistically by simultaneous introduction of two equimolar dopants~\cite{Liu2016,Wen2017,QWen2017,YSun2017,Liu2019,BFeng2018,SDas2019,BFeng2019,FLi2019,Li2019,ZhouLi2019,He2020,Kim2021,Yin2021}. 

Despite promising thermoelectric properties of BiCuSeO, a fundamental understanding of its physical properties is still lacking. In particular, the unusually low lattice thermal conductivity has many proposed theoretical explanations~\cite{Saha2015,Ding2015,Shao2016,GLiu2016,Kumar2016,Ji2016,SKSaha2016,DDFan2017,Chang2018,TZhao2020}, but direct experimental test of these theories has been sparse and inconclusive~\cite{Vaqueiro2015,Viennois2019,Lin2020}. Moreover, while the role of Ca doping in reducing the lattice thermal conductivity in BiCuSeO is well understood, with Ca forming calcium oxide nanoparticles that enhance phonon scattering~\cite{Pei2013,He2020}, the underlying mechanism of silver (Ag) doping in reducing lattice thermal conductivity in single and dual-doped BiCuSeO is unclear~\cite{Tan2014,LiuAg2015,Farooq2017,Li2019}. 

This work presents a systematic study of undoped, Ag doped and (Pb,Ag)$-$dual doped BiCuSeO, with the aim to examine the origin of the intrinsic low lattice thermal conductivity, and mechanism of further reduction by introduction of Ag. For this, the experimental techniques of powder X-ray diffraction (XRD) and Extended X-ray Absorption Fine Structure (EXAFS) are employed to understand the global and local crystal structure. Temperature dependent XRD and EXAFS measurements are carried out, to investigate the dynamics through thermal displacement parameters, in order to directly study the origins of the low lattice thermal conductivity in BiCuSeO. 

While XRD provides accurate results about the overall `global' crystal structure, it is not sensitive to dopants with small concentrations. In order to study structure and dynamics of these dopants, EXAFS measurements are conducted. EXAFS is an element specific spectroscopic technique, that can be used to gain information about the doping site of trace elements in a crystal structure, their local environment and motion along different atom-pair directions. Thus, the combination of XRD and EXAFS provides a complete picture of the global crystal structure and local sites of dopants in this structure, along with information about thermal dynamics of the atoms.

\section{\label{sec:methods}Experimental methods}

\subsection{Sample Preparation}

Polycrystalline (powder) samples of undoped BiCuSeO, Ag doped BiCuSeO (BiCu$_{1-x}$Ag$_x$SeO) and (Pb,Ag)$-$dual doped BiCuSeO (Bi$_{1-x}$Pb$_x$Cu$_{1-x}$Ag$_x$SeO) for doping levels of $x=\{0.02,0.04,0.06,0.08\}$ were synthesized via a solid state reaction process. The raw materials used were Bi (Alfa Aesar, 99.5\%), Bi$_2$O$_3$ (Alfa Aesar, 99.9\%), Cu (Alfa Aesar, 99.9\%), PbO (Sigma Aldrich, 99.9\%), Ag (Sigma Aldrich, 99.9\%) and Se (Alfa Aesar, 99.5\%) powders. Stoichiometric mixtures of these raw materials were mixed and ground in an agate mortar and pestle. The mixtures were pressed into cylindrical pellets of 1.6 cm diameter under 50 MPa. The pellets were sealed in quartz tubes under vacuum ($<$10$^{-2}$ mbar). The sealed tubes were put in a box furnace, heated to 623 K and held there for 12 h. This intermediate step ensures complete reaction of Se with the elemental metals~\cite{Tan2014}, and reduction in impurity phases, like Bi metal, in the final synthesized sample~\cite{Hiramatsu2008}. Then, the temperature of the furnace was raised to 973 K and held for 48 h. The tubes were then furnace-cooled to room temperature. The obtained samples were ground, re-pelletized and reheated in sealed quartz tubes at 973 K for another 48 h to improve purity and crystallinity. All heating sequences were carried out at a ramp rate of 100 K/h. Finally, dark gray powder samples were obtained, which were found to have greater than 99.5\% phase purity, through XRD analysis. The primary impurity phase was unreacted Bi$_2$O$_3$, as seen in Figure~\ref{fig:XRD2}, which is a known limitation of the solid state reaction process~\cite{Ren2015}. Details of phase purity analysis are included in the Supplemental Material~\cite{[{See Supplemental Material at }]supp}. 

\subsection{X-ray Diffraction}

All XRD measurements were done using a Rigaku Smartlab diffractometer. The powder samples were thoroughly ground before the measurement, and mounted in Bragg-Brentano reflection geometry. Cu K$\alpha$ radiation was used, with Cu K$\beta$ blocked using a nickel filter. Heating and cooling from room temperature was done using Anton Paar DHS and DCS sample stages respectively, under vacuum ($<$10$^{-1}$ mbar). The DHS stage was used in the temperature range of 298$-$1023 K, while the DCS stage was used in the 93$-$773 K range. In the overlap temperature range of 303$-$773 K, all the XRD data and fitted parameters were found to agree, within measurement uncertainty. Hence, for visual clarity, data from the DCS stage is only shown for 93$-$298 K range, with the DHS stage data displayed for the range of 298$-$1023 K. Note that the measurements were only done up to 1023 K since BiCuSeO decomposes at higher temperatures~\cite{LiJ2014, BarreteauJ2015, Sato2016}, which is intuitively expected given the synthesis temperature of BiCuSeO. 

Rietveld refinement of XRD data of all samples was done on the GSAS-II software~\cite{Toby2013}. Reitveld refinement produced fits for the data and background, and yielded the lattice constants (\textit{a}, \textit{c}) of the tetragonal unit cell and isotropic mean-square displacements ($U_{iso}$) for each atom. Note that $U_{iso}$ was computed to obtain values with reasonable error bars, since the refinement of anisotropic displacement parameters with X-ray diffraction increases uncertainty significantly. To overcome this difficulty, we used EXAFS to estimate atom-pair mean-square displacement, as described in the next section. 

\subsection{EXAFS}
All EXAFS measurements were carried out at the Canadian Light Source (CLS) on the Hard X-ray Micro-Analysis (HXMA) beamline 06ID-1. At room temperature (298 K), Ag K-edge (25514 eV) data was collected for all doped samples, while Cu K-edge (8979 eV) data was collected for all doped and undoped samples. For temperature dependence, the highest doping concentration ($x=0.08$) of single doped and dual-doped BiCuSeO samples were chosen, along with the undoped BiCuSeO sample. The energy of the incident X-ray beam from the wiggler was selected using a double-crystal Si(111) monochromator, with higher harmonic contributions suppressed by a combination of Rh-coated mirrors and a 50\% detuning of the wiggler. Relatively low concentration of Ag made transmission geometry unsuitable. Ag K-edge EXAFS measurements were performed in a 90 degree, fluorescence geometry using a 13-element Ge solid-state detector, with the sample 45 degree to the incident beam. Cu K-edge EXAFS data, on the other hand, was collected in both transmission and fluorescence geometry simultaneously, with the sample again 45 deg to the incident beam. Standard Ag and Cu foils were utilized for energy calibration, respectively. As reference, EXAFS spectra in transmission geometry were also collected on these standard foils, along with each EXAFS measurement on the actual samples. Each sample was diluted with pure boron nitride powder, finely ground, pressed into a uniform pellet, and then affixed to polyimide tape. These procedures were done in order to achieve suitable element concentration and uniformity. At least three scans were taken at each edge for each sample in order to guarantee data reproducibility and good signal-to-noise ratio. 

EXAFS data processing and fitting was done on the ATHENA and ARTEMIS softwares of the Demeter suite respectively~\cite{Ravel2005}. In data processing, EXAFS data obtained in energy-space was reduced using standard procedures, and the resulting $k$-space data was Fourier transformed (F.T.) to $r$-space, yielding peaks that correspond to different shells of neighboring atoms in the EXAFS equation:\\
\begin{equation}\label{eq:eq1}
\chi(k) = \sum\limits_j \frac{S_0^2 N_j e^{-2k^2\sigma^2_j}e^{-2r_j/\lambda(k)}f_j(k)}{kr_j^2}\sin(2kr_j + \delta_j(k))
\end{equation}~\\
where $j$ represents different coordination shells made up of $N_j$ identical atoms at approximately the same distance from the absorbing atom; $S_0^2$ is the amplitude reduction factor which corrects for multi$-$electron scattering and other experimental effects (generally between 0.7 and 1.2); $\sigma^2_j$ is the atom-pair mean-square displacement along the atom-pair distance $r_j$; $f_j(k)$ and $\delta_j(k)$ are the scattering factor and scattering phase shift, which are provided in the FEFF program~\cite{Rehr1995}; $\lambda(k)$ is the mean-free-path of the photoelectron; $k = \sqrt{2m(E-E_0)/\hbar^2}$, where $E_0$ is the theoretical Fermi level position. In data fitting, the $r$-space data is fit to a sum of complex phase and amplitude functions calculated using the FEFF program, to extract atom-pair distances ($r_j$) and thermal factors ($\sigma^2_j$) for each atom-pair. Another two fitting parameters, $S_0^2$ and $\Delta E_0$ (the difference in edge energy between the value defined for the data and the theoretical function), are also determined for each edge. In the preliminary fits, the values of $S_0^2$ are found to be reasonable, very close to 1, for all temperatures. Hence, in the final fits, $S_0^2$ is fixed to its room temperature value for all edges. In this analysis, we fit both the real and imaginary parts of the $r$-space data. 

\begin{table*}[htbp]
\caption{\label{tab:table1}Lattice constants \textit{a} and \textit{c} of the tetragonal unit cell, with the space group P4/nmm (No. 129), for BiCu$_{1-x}$Ag$_x$SeO and Bi$_{1-x}$Pb$_x$Cu$_{1-x}$Ag$_x$SeO samples, obtained from Rietveld refinement of XRD data taken at room temperature (298 K).}
\begin{ruledtabular}
\begin{tabular}{ccccc}
Sample&\multicolumn{2}{c}{BiCu$_{1-x}$Ag$_x$SeO}&\multicolumn{2}{c}{Bi$_{1-x}$Pb$_x$Cu$_{1-x}$Ag$_x$SeO}\\
x & $a$ [\textup{\AA}] & $c$ [\textup{\AA}] & $a$ [\textup{\AA}] & $c$ [\textup{\AA}]\\ \hline
0.00 (undoped) & 3.928(4) & 8.931(9) & 3.928(4) & 8.931(9)\\
0.02 & 3.929(4) & 8.932(9) & 3.930(4) & 8.947(9)\\
0.04 & 3.929(4) & 8.932(9) & 3.932(4) & 8.956(9)\\
0.06 & 3.930(4) & 8.933(9) & 3.935(4) & 8.976(9)\\
0.08 & 3.930(4) & 8.934(9) & 3.935(4) & 8.984(9)\\
\end{tabular}
\end{ruledtabular}
\end{table*}

\section{\label{sec:results}Experimental results}

\subsection{X-ray Diffraction}

\subsubsection{Room Temperature}

The XRD data obtained at room temperature for all samples was fit using Rietveld refinement to extract the lattice constants of the tetragonal unit cell. The lattice constants obtained from this fitting are tabulated in Table~\ref{tab:table1}. It is worth noting that, in the case of Ag doping, the lattice constants show almost no change as doping increases, within error bar. In contrast, in the case of (Pb,Ag)$-$dual doping, the lattice constants show a more pronounced increase with doping, in particular the \textit{c}-lattice constant. 

Isotropic mean-square displacements ($U_{iso}$) were also obtained for each atom from Rietveld refinement of XRD data of undoped BiCuSeO, which will be discussed in the next section. Due to the relatively low concentration of dopants, $U_{iso}$ values for dopant atoms could not be estimated by Rietveld refinement. 

\begin{figure*}[htbp]
\centering
\begin{subfigure}[b]{0.45\textwidth}
\includegraphics[width=\linewidth]{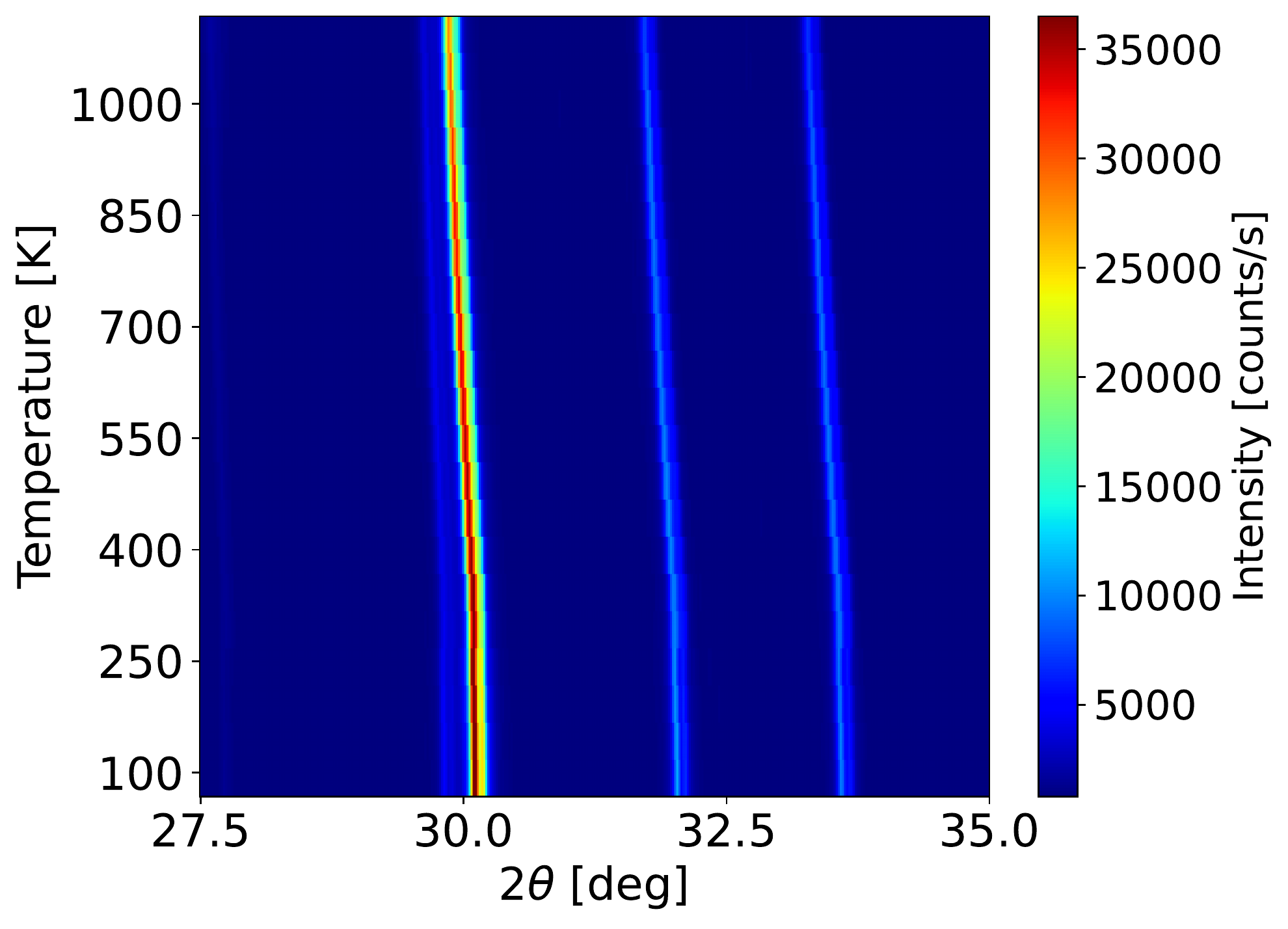}
\caption{\label{fig:XRD1}}
\end{subfigure}
\begin{subfigure}[b]{0.5\textwidth}
\includegraphics[width=\linewidth]{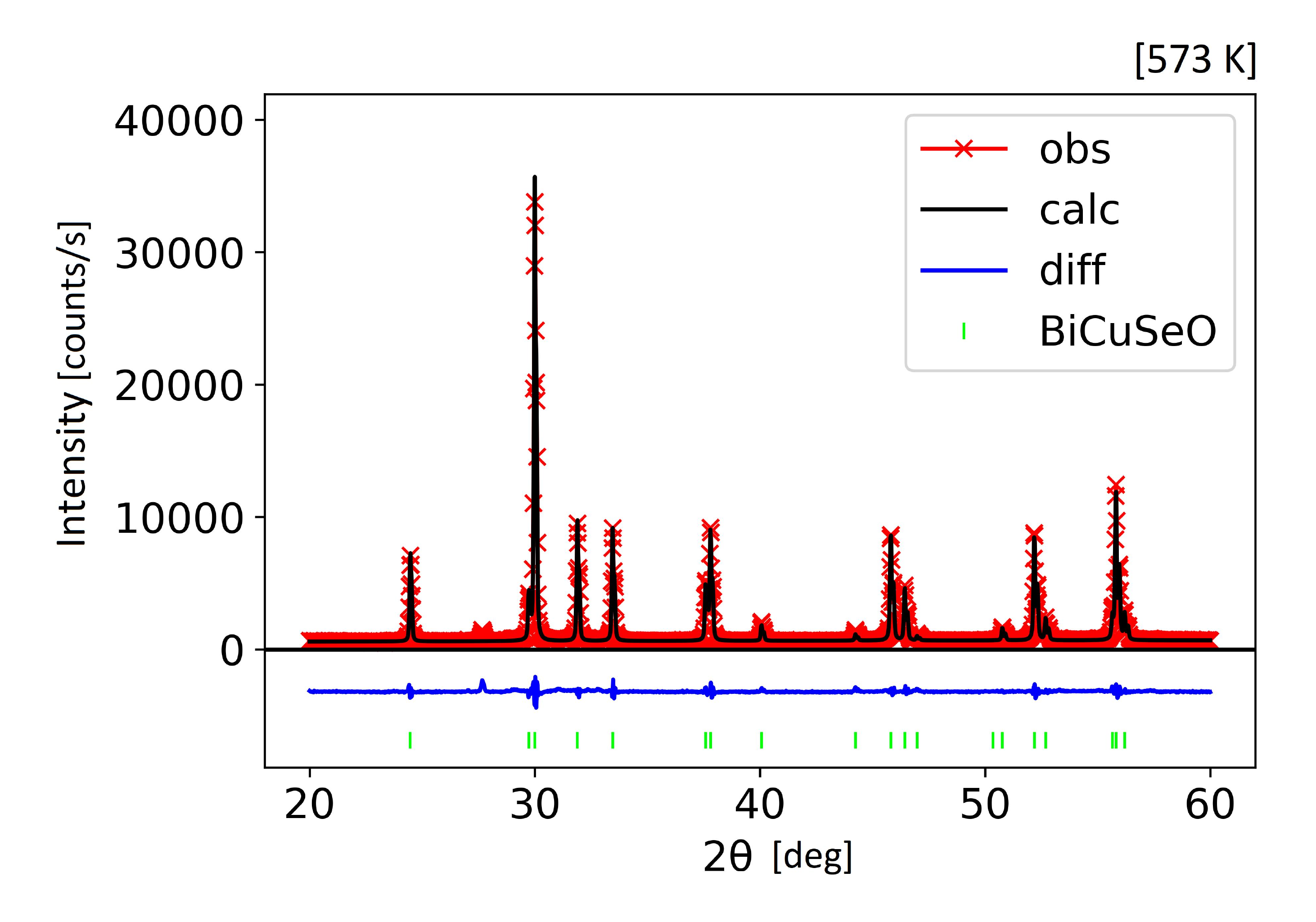}
\caption{\label{fig:XRD2}}
\end{subfigure}
\caption{\subref{fig:XRD1}~XRD raw data for undoped BiCuSeO, as a function of temperature, zoomed in near the most prominent XRD peak at 30 deg. Thermal expansion is clearly observed, depicted by the decrease in 2$\mathrm{\theta}$ position of peaks with increase in temperature. \subref{fig:XRD2}~Rietveld refinement profile of the XRD data for undoped BiCuSeO at 573 K. The calculated profile, shown in black, is seen to fit the observed data well, shown in red. The difference profile and expected XRD peak positions are also shown in blue and green respectively. The primary impurity XRD peak of unreacted Bi$_2$O$_3$ (less than 0.5\%) is seen near 27 deg.}
\end{figure*}

\subsubsection{Temperature Dependence}

XRD data obtained for undoped BiCuSeO as a function of temperature is shown in Figure~\ref{fig:XRD1}. The color-map shows intensity of XRD data as a function of temperature. Only the limited 2$\theta$ range of 27.5$-$35 deg is shown to illustrate the temperature dependence clearly. The colored lines of higher intensity represent the XRD peak positions from the sample. Note that as temperature increases, the peak position shifts to lower angles, resulting from an increase in lattice constants due to thermal expansion. No discontinuity is observed in the data over the entire temperature range, signifying that no phase change occurs in this range. 

The Rietveld refinement profile for XRD data of undoped BiCuSeO at 573 K is shown in Figure~\ref{fig:XRD2}, for the 2$\theta$ range of 20$-$60 deg. The observed data is shown in red, while the calculated fit is the solid black line. The fit is seen to match the data well, yielding R$_{\mathrm{wp}}=5.33\%$ and $\chi^2_r=2.19$. These goodness of fit parameters are found to be similar in value for fits at all temperatures. The difference profile is shown in blue, and is found to be within the measurement uncertainty throughout the entire range. The expected XRD peak positions for BiCuSeO, based on the crystal structure in Figure~\ref{fig:fig1_structure}, are shown via the green vertical lines. 

\begin{figure*}[htbp]
\centering
\begin{subfigure}[b]{0.48\textwidth}
\includegraphics[width=\linewidth]{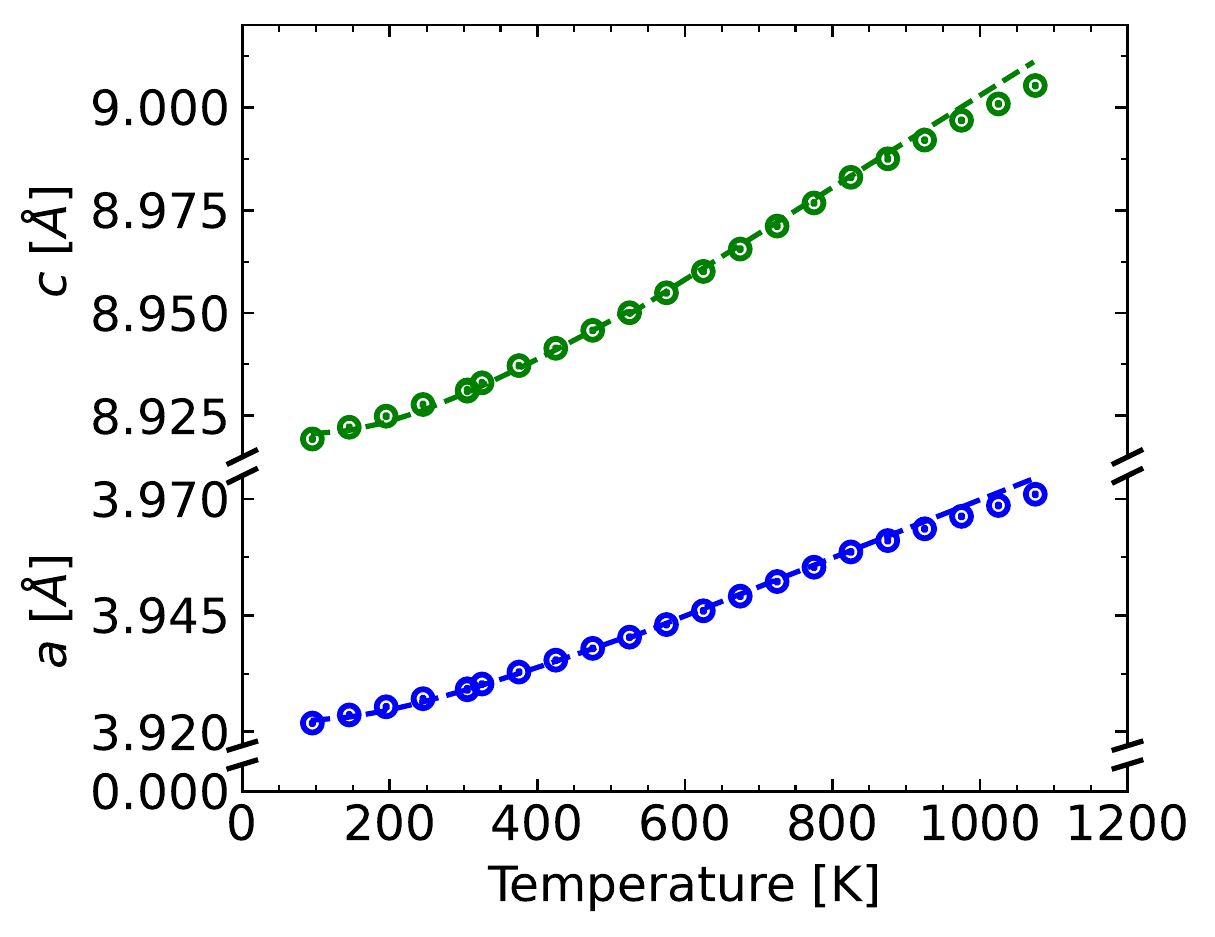}
\caption{\label{fig:LC}}
\end{subfigure}
\begin{subfigure}[b]{0.5\textwidth}
\includegraphics[width=\linewidth]{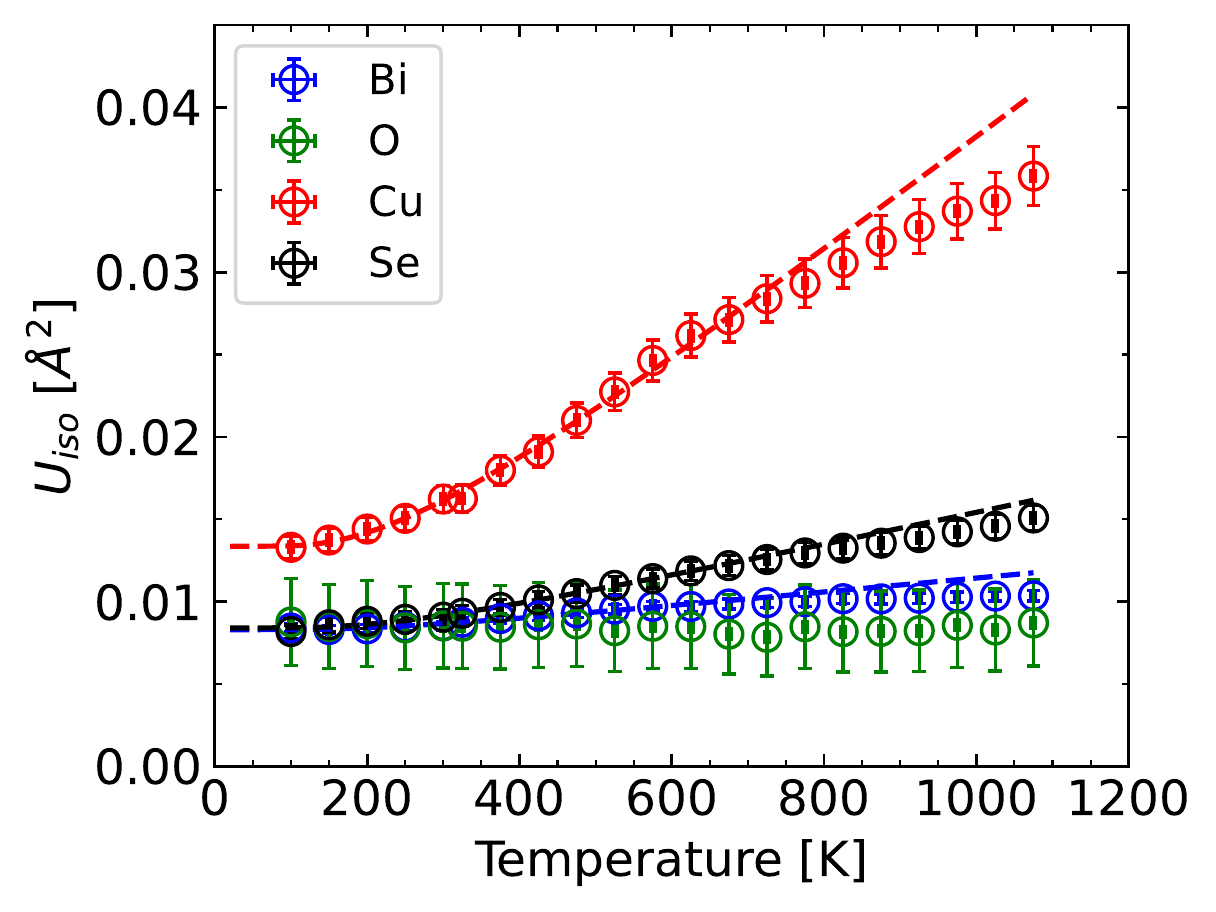}
\caption{\label{fig:SS}}
\end{subfigure}
\caption{\label{fig:lattice}\subref{fig:LC}~Temperature dependence of lattice constant \textit{a} and lattice constant \textit{c} of the tetragonal unit cell for undoped BiCuSeO. The dashed lines represent quasi-harmonic fits up to 800 K, described by Eq.~\eqref{eq:eq2}, based on the Einstein model. \subref{fig:SS}~Isotropic mean-square displacements ($U_{iso}$) versus temperature for Bi, Cu, O and Se in undoped BiCuSeO. The dashed lines represent quasi-harmonic fits for the mean-square displacements up to 700 K. Both lattice constants and isotropic mean-square displacements were obtained from Rietveld refinement of XRD data of undoped BiCuSeO.}
\end{figure*}

Rietveld refinement of the temperature dependent XRD data yields the temperature dependence of the lattice constants of the tetragonal unit cell, and isotropic mean-square displacements of each atom, for undoped BiCuSeO, which is presented in Figure~\ref{fig:lattice}. The temperature dependence of the lattice constant \textit{a} is shown in the lower half of Figure~\ref{fig:LC}. Uniform thermal expansion is observed from base temperature (93 K) to room temperature (298 K). Above room temperature, there is a subtle increase in the rate of expansion up to 700 K, following which the rate decreases up to 1023 K. The relative thermal expansion behavior is almost identical for lattice constant \textit{c}, shown in the upper half of Figure~\ref{fig:LC}. The error bars for both lattice constants are smaller than the symbol sizes shown in the plot. 

The Einstein model is often used as a quasi-harmonic fit for thermal expansion of lattice constants~\cite{Mi2011,Jorgensen2001}: \\
\begin{equation}\label{eq:eq2}
\ln\left(\frac{a}{a_0}\right) = \frac{A\; \Theta_E}{\exp(\Theta_E/T) - 1}
\end{equation}~\\
where $a$ is the length of the lattice constant at temperature $T$, $a_0$ is the lattice constant at $T$ = 0 K, $A$ is a scaling constant and $\Theta_E$ is the Einstein temperature. A similar relation is used to fit lattice constant $c$. The quasi-harmonic fits, based on Eq.~\eqref{eq:eq2}, are shown via dashed lines in Figure~\ref{fig:LC}. These quasi-harmonic fits match the data for the temperature range of 93$-$800 K, after which the data for both lattice constants diverges subtly from the fits. The relative behaviour of lattice constants is found to be similar to doped samples. The value of $\Theta_E$ is found to be 95 ($\pm$1) K, which matches the value quoted in previous literature~\cite{Vaqueiro2015}, within experimental uncertainty. $\Theta_E$ is also consistent with the value obtained through harmonic fits of $U_{iso}$, as discussed below. 

Isotropic mean-square displacements ($U_{iso}$) of Bi, Cu, O and Se are shown in Figure~\ref{fig:SS}. Note that $U_{iso}$ for oxygen shows large uncertainty, due to the small atomic form factor of oxygen. The dashed lines correspond to the quasi-harmonic fits for the temperature dependence of $U_{iso}$ for each element. The quasi-harmonic fits only work for the temperature range of 93$-$700 K, after which the data for all three elements is clearly seen to diverge significantly from the harmonic fits. Both Einstein model and Debye model were used as quasi-harmonic fits for the data~\cite{Bentien2005,Mi2011}:
\begin{eqnarray}\label{eq:eq3}
&\mathrm{Einstein}: U_{iso} = \frac{\hbar^2}{2mk_B\Theta_E}\coth\left(\frac{\Theta_E}{2T}\right) + d_E^2 \\
&\mathrm{Debye}: U_{iso} = \frac{3\hbar^2T}{mk_B\Theta_D^2}\left[\frac{T}{\Theta_D}\int\limits_0^{\Theta_D/T} \frac{x}{e^x - 1}dx + \frac{\Theta_D}{4T}\right] + d_D^2
\end{eqnarray}
where $\hbar$ is the reduced Planck constant, $m$ is the mass of the atom, $k_B$ is the Boltzmann constant, $\Theta_E$ is the Einstein temperature, $\Theta_D$ is the Debye temperature, $T$ is the sample temperature and $d_E$ \& $d_D$ are empirical terms to describe temperature-independent disorder. The two fits were found to overlap almost completely, with almost same $d_E$ and $d_D$. Hence, only the Einstein fits are shown for visual clarity. 

The value of $\Theta_E$ and $\Theta_D$ from the fits are 95 ($\pm$2) K and 241 ($\pm$3) K respectively, which matches values reported previously by other studies~\cite{Vaqueiro2015,Zhao2014}. It is interesting to note that these values are also very similar to the values obtained for other thermoelectric materials like clathrates~\cite{Bentien2005} and skutterudites~\cite{Mi2011}. 

Note that the values of $U_{iso}$ are significantly higher for Cu, as compared to Bi, O and Se, suggesting that the displacement amplitude for Cu is higher than the other elements. In general, tightly bound atoms in a solid crystal structure are expected to have $U_{iso}$ vales close to 0.007 \AA$^2$ at room temperature~\cite{Toby2013,ch5}. At room temperature, while Bi, Se and O values match this expected value within error bar, the $U_{iso}$ value for Cu is 0.016 \AA$^2$. Also note that $d_E^2$ for Cu is much higher than the other elements. This indicates that disorder on Cu is significantly larger than that for other elements in BiCuSeO. Our results is consistent with a previous neutron diffraction study~\cite{Vaqueiro2015}, confirming the large thermal parameters of Cu over a wide temperature range, from low temperatures up to the decomposition temperature of undoped BiCuSeO. 

\begin{figure*}[htbp]
\centering
\begin{subfigure}[b]{0.45\textwidth}
\includegraphics[width=\linewidth]{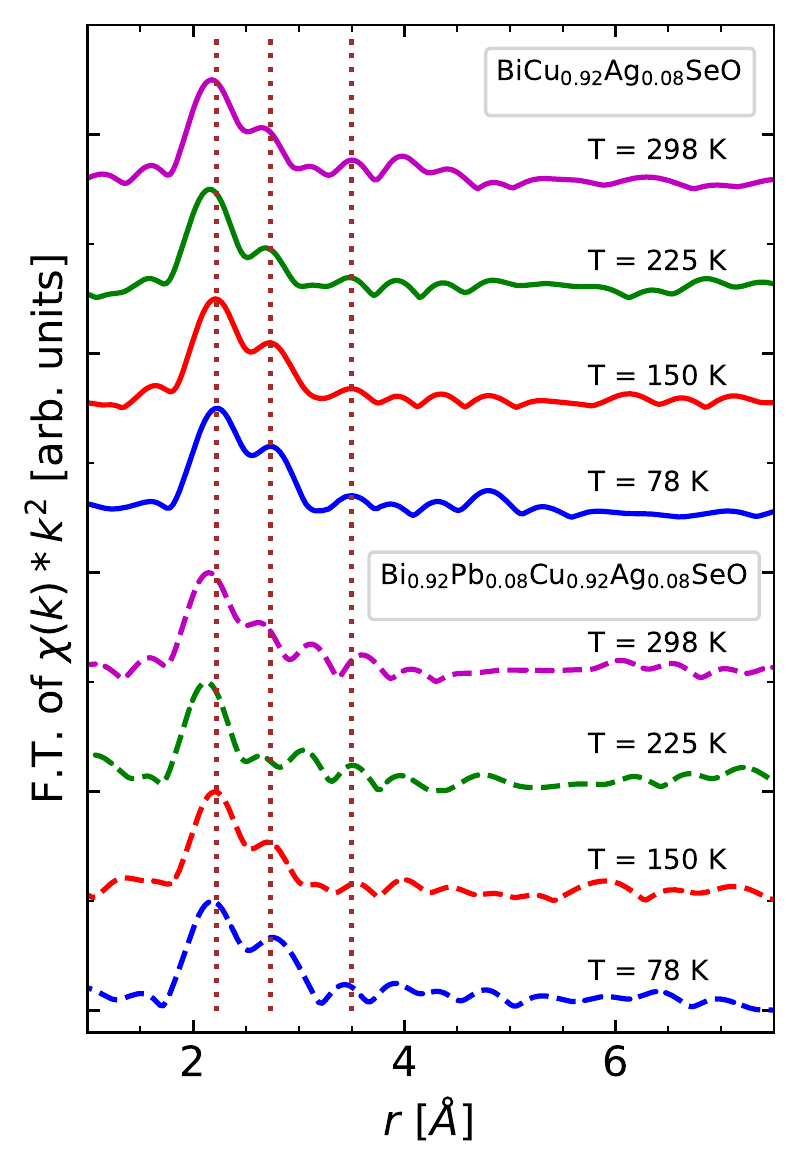}
\caption{\label{fig:Ag1}}
\end{subfigure}
\begin{subfigure}[b]{0.45\textwidth}
\includegraphics[width=\linewidth]{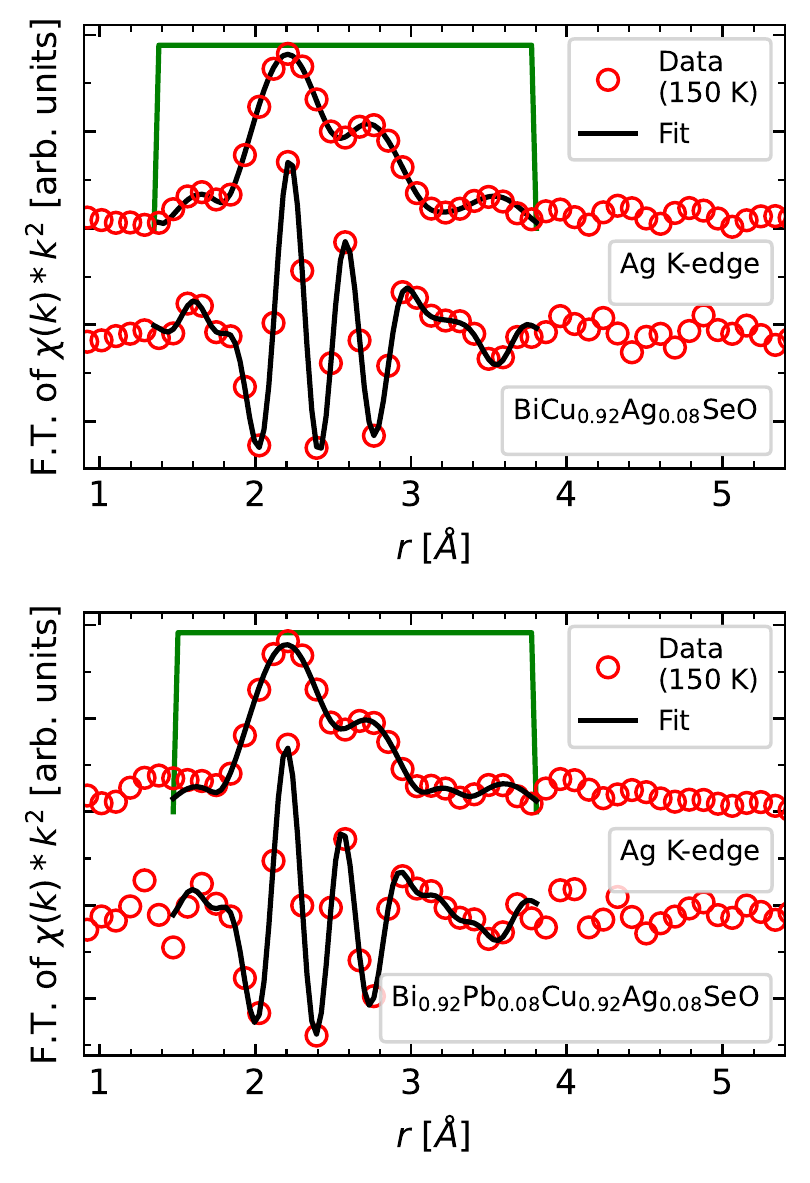}
\caption{\label{fig:Ag2}}
\end{subfigure}
\caption{\subref{fig:Ag1}~Ag K-edge EXAFS data in $r$-space for BiCu$_{0.92}$Ag$_{0.08}$SeO as solid lines, and Bi$_{0.92}$Pb$_{0.08}$Cu$_{0.92}$Ag$_{0.08}$SeO as dashed lines, at different temperatures. The vertical brown dotted lines represent the first (Ag-Se), second (Ag-(Cu/Ag)) and third (Ag-(Bi/Pb)) shell locations in $r$-space, corresponding to Ag replacing Cu in the BiCuSeO crystal structure framework. The Fourier transform (F.T.) ranges are 3$-$11 \textup{\AA}$^{-1}$. \subref{fig:Ag2}~The fit results for BiCu$_{0.92}$Ag$_{0.08}$SeO and Bi$_{0.92}$Pb$_{0.08}$Cu$_{0.92}$Ag$_{0.08}$SeO at 150 K. The fit ranges are shown via the window function in solid green lines. The oscillatory curve below the envelope is the real part of the Fourier transform ($FT_R$). The envelope is the magnitude of the Fourier transform, defined as $\sqrt{FT_R^2 + FT_I^2}$, where $FT_I$ (not shown) is the imaginary part of the Fourier transform.}
\end{figure*}

\begin{table*}[htbp]
\caption{\label{tab:table2}Mean-square displacement ($\sigma^2_j$) along atom-pairs of Ag-Se ($\sigma^2_1$), Ag-(Cu/Ag) ($\sigma^2_2$) and Ag-(Bi/Pb) ($\sigma^2_3$) for Ag doped BiCuSeO and (Pb,Ag)$-$dual doped BiCuSeO samples at different temperatures, obtained from fitting of Ag K-edge EXAFS data.}
\begin{ruledtabular}
\begin{tabular}{ccccccc}
Sample&\multicolumn{3}{c}{BiCu$_{0.92}$Ag$_{0.08}$SeO}&\multicolumn{3}{c}{Bi$_{0.92}$Pb$_{0.08}$Cu$_{0.92}$Ag$_{0.08}$SeO}\\
Temperature [K] & $\sigma^2_1$ [\textup{\AA}$^2$] & $\sigma^2_2$ [\textup{\AA}$^2$] & $\sigma^2_3$ [\textup{\AA}$^2$] & $\sigma^2_1$  [\textup{\AA}$^2$] & $\sigma^2_2$ [\textup{\AA}$^2$] & $\sigma^2_3$ [\textup{\AA}$^2$]\\ \hline
80 & 0.0005(1) & 0.0107(5) & 0.0003(1) & 0.0008(6) & 0.0107(8) & 0.0007(4)\\
150 & 0.0006(2) & 0.0103(7) & 0.0007(3) & 0.0006(5) & 0.0107(4) & 0.0004(1)\\
225 & 0.0007(6) & 0.0107(7) & 0.0007(6) & 0.0009(5) & 0.0106(6) & 0.0004(2)\\
298 & 0.0008(2) & 0.0106(5) & 0.0002(1) & 0.0006(2) & 0.0108(3) & 0.0005(1)\\
\end{tabular}
\end{ruledtabular}
\end{table*}

\subsection{EXAFS}

\subsubsection{Ag K-edge}

Having characterized the temperature dependence for undoped BiCuSeO, the doped samples are now characterized using temperature dependent Ag K-edge EXAFS measurements. The samples showed no significant doping dependence for room temperature EXAFS measurements, which means that the local environment of the dopant site remains the same as doping fraction increases. Further information about doping dependence can be found in the Supplemental Material~\cite{[{See Supplemental Material at }]supp}. Hence, the highest doping percentage sample ($x=0.08$) was chosen for both single and dual-doped BiCuSeO for temperature dependent EXAFS, since higher concentration leads to better signal-to-noise ratio. 

Figure~\ref{fig:Ag1} shows the Ag K-edge EXAFS data (magnitude of Fourier transform) in $r$-space as a function of temperature. The data for the single doped sample is shown with solid lines in the upper half, while the dashed lines in the lower half correspond to data for the dual-doped sample. The vertical dotted lines are shown as visual aids to note the positions of the first three shells expected in the EXAFS data for Ag replacing Cu in BiCuSeO, based on the structure shown in Figure~\ref{fig:fig1_structure}. These shells correspond to the Ag-Se ($r$ = 2.2 \AA), Ag-(Cu/Ag) (2.7 \AA) and Ag-(Bi/Pb) (3.5 \AA) atom-pairs respectively. Note that the $r$-space peak positions of shells are different from the exact atom-pair distances, which are tabulated in the Supplemental Material~\cite{[{See Supplemental Material at }]supp}. 

No significant temperature dependence is observed in this data across both samples. It is generally expected that low temperature ($T$$\ll$$\Theta_D$) EXAFS data has better signal-to-noise ratio, due to smaller Debye-Waller factor. However, this temperature effect of the Debye-Waller factor is due to a reduction in the mean-square displacements upon cooling. Thus, the lack of temperature dependence indicates that the mean-square displacement of Ag dopant does not change significantly in the temperature range studied, from room temperature (298 K) down to base temperature (78 K). This mean-square displacement behavior of Ag for this temperature range is very similar to that of Cu, based on $U_{iso}$ from the XRD data shown in Figure~\ref{fig:SS}.

To obtain quantitative information, the EXAFS data is fit to the first three shells expected for Ag replacing Cu in BiCuSeO. This three-shell fit for the data at 150 K is shown in Figure~\ref{fig:Ag2}. The upper panel corresponds to the single doped sample while the lower panel displays the fit for the dual-doped sample. The fitting process involves fitting both the magnitude ($\sqrt{FT_R^2 + FT_I^2}$) and real part ($FT_R$) of the Fourier transform. In Figure~\ref{fig:Ag2}, the oscillatory curve is the real part, while the envelope is the magnitude of the Fourier transform. The fits match the data well in the fitting window, yielding a goodness of fit parameter, reduced-$\chi^2$, close to 1 for all temperatures. 

The fits yield the atom-pair mean-square displacements ($\sigma^2_j$) for the three fitted shells, tabulated in Table~\ref{tab:table2}. The values of $\sigma^2_j$ for every shell are found to be almost temperature independent for both samples. This was already expected, based on the lack of temperature dependence seen in the raw data in Figure~\ref{fig:Ag1}. The values of $\sigma^2_2$, corresponding to Ag-(Cu/Ag) atom-pair, are an order of magnitude higher than the values of $\sigma^2_j$ for the other atom-pairs. These values are slightly smaller than the $U_{iso}$ of Cu at 150 K obtained by XRD, and significantly larger than the $U_{iso}$ values of Bi and Se at the same temperature, based on the $U_{iso}$ plot shown in Figure~\ref{fig:SS}. This suggests that the significant disorder on Cu in the BiCuSeO crystal structure framework is retained by the Ag dopant, on replacing Cu at that site, for all temperatures studied. This means that the disorder observed is actually linked to the Cu-site, as opposed to just the Cu atom, since Ag dopant retains the observed disorder. 

Note that the comparison of $U_{iso}$ and $\sigma^2_j$ is possible since both represent different measures of the same physical quantity, mean-square displacement. Since both account for thermal vibrations of atoms about their equilibrium positions, it is expected that all contributions from different atom-pairs in the various $\sigma^2_j$ add up to give $U_{iso}$~\cite{Beni1976}. Thus, the largest $\sigma^2_j$ is expected to influence $U_{iso}$ the most. In the case of Ag K-edge EXAFS, it is found to be the second shell, Ag-(Cu/Ag). 

\begin{table*}[htbp]
\caption{\label{tab:table3}Mean-square displacement ($\sigma^2_j$) along atom-pairs of Cu-Se ($\sigma^2_1$) and Cu-(Cu/Ag) ($\sigma^2_2$) for undoped BiCuSeO,  Ag doped BiCuSeO (BiCu$_{0.92}$Ag$_{0.08}$SeO) and (Pb,Ag)$-$dual doped BiCuSeO (Bi$_{0.92}$Pb$_{0.08}$Cu$_{0.92}$Ag$_{0.08}$SeO) samples at different temperatures, obtained from fitting of Cu K-edge EXAFS data.}
\begin{ruledtabular}
\begin{tabular}{ccccccc}
Sample&\multicolumn{2}{c}{BiCuSeO}&\multicolumn{2}{c}{BiCu$_{0.92}$Ag$_{0.08}$SeO}&\multicolumn{2}{c}{Bi$_{0.92}$Pb$_{0.08}$Cu$_{0.92}$Ag$_{0.08}$SeO}\\
Temperature [K] & $\sigma^2_1$ [\textup{\AA}$^2$] & $\sigma^2_2$ [\textup{\AA}$^2$] & $\sigma^2_1$ [\textup{\AA}$^2$] & $\sigma^2_2$  [\textup{\AA}$^2$] & $\sigma^2_1$ [\textup{\AA}$^2$] & $\sigma^2_2$ [\textup{\AA}$^2$]\\ \hline
150 & 0.0012(1) & 0.013(2) & 0.0013(2) & 0.010(1) & 0.0013(3) & 0.013(2)\\
225 & 0.0011(3) & 0.011(5) & 0.0010(4) & 0.012(4) & 0.0012(2) & 0.012(3)\\
298 & 0.0010(5) & 0.013(3) & 0.0012(2) & 0.011(3) & 0.0013(1) & 0.011(2)\\
\end{tabular}
\end{ruledtabular}
\end{table*}

\subsubsection{Cu K-edge}

Since Ag is expected to replace Cu in BiCuSeO, Cu K-edge EXAFS measurements were also performed in order to provide a reference point of the same site. Also, having drawn the comparison of similarities in mean-square displacement values of Ag (from EXAFS) and Cu (from XRD), these measurements provide a direct comparison between the two elements based on the same technique of temperature dependent EXAFS. These measurements were conducted on the undoped, 8\% Ag doped and 8\% (Pb,Ag)$-$dual doped samples. 

The fit results for Cu K-edge EXAFS measurements are summarized in Table~\ref{tab:table3}. The atom-pair mean-square displacement values for the first two shells, $\sigma^2_1$ (Cu-Se) and $\sigma^2_2$ (Cu-(Cu/Ag)) are shown as a function of temperature. Note that higher noise levels of lower energy Cu K-edge, coupled with significant disorder, limited the fitting range for the EXAFS data, hence only two shell fits could be estimated with reasonable error bars. No significant temperature dependence is observed, as expected for this temperature range, based on the $U_{iso}$ plot for Cu, shown in Figure~\ref{fig:SS}. 

The values of $\sigma^2_2$, corresponding to Cu-(Cu/Ag) atom-pair, are an order of magnitude higher than the values of $\sigma^2_j$ for the other atom-pair. These values agree well with the $U_{iso}$ of Cu at 150 K obtained by XRD, shown in Figure~\ref{fig:SS}. Our combined XRD and EXAFS results indicate that the displacement parameters for Cu do not change significantly below room temperature. Note that this observation disagrees with an earlier neutron diffraction study on undoped BiCuSeO, in which a fairly strong temperature dependence of anisotropic atomic displacement parameters (ADPs) were reported~\cite{Viennois2019}. We also note that we decided not to estimate anisotropic parameters in our studies due to the large uncertainty associated with estimating these parameters from diffraction data.

The Cu K-edge EXAFS results confirm the observation of significant disorder at the Cu-site in both undoped and doped BiCuSeO, which arises mainly along the Cu-(Cu/Ag) atom-pair. Note that the introduction of Ag at Cu-sites in the doped samples does not affect the disorder of Cu atoms, since the values of $\sigma^2_2$ are almost same for all dopings, at all temperatures. The values of overall mean-square displacement for Cu from EXAFS (sum of $\sigma^2_j$) and XRD ($U_{iso}$) are found to match within error bar, suggesting that contributions from further shells of neighbouring atoms are not significant, compared to the second shell, Cu-(Cu/Ag), in particular. 

\section{\label{sec:discuss}Discussion}

The main findings from our X-ray diffraction and EXAFS studies are our observation of the unusually large $U_{iso}$ parameters for Cu in undoped BiCuSeO and large $\sigma^2$ parameters for both Ag and Cu for doped BiCuSeO samples. We also observe that these ``thermal” displacement parameters are independent of doping concentrations as well as temperature (at low temperatures). We will first discuss the possible origin of the observed large “thermal” parameters and the implication of our observations for thermoelectric properties of doped and undoped BiCuSeO.

First, we note that $U_{iso}$ parameters as well as $\sigma_j^2$ parameters are usually called thermal parameters because the main source of this displacement parameter is due to lattice dynamics (i.e. Debye-Waller factor), but static disorder can also contribute to these parameters. Often dynamic and static origins of the disorder can be distinguished from the temperature dependence of thermal parameters, since the dynamic contribution shows characteristic temperature dependence. However, temperature-independent thermal parameter does not automatically indicate static disorder, since quasi-static disorder, such as a low frequency fluctuation, will be temperature independent for most of the temperature range probed.

While our observed large thermal parameters for Cu and Ag in Cu-site (this is further discussed below) could be either due to static or quasi-static disorder, we argue that static disorder is less likely to be the main source of the disorder. What is peculiar about the $U_{iso}$ data shown in Figure~\ref{fig:SS} is that the lowest temperature $U_{iso}$ value of Cu is at least 50\% larger than other elements in BiCuSeO. This is surprising because we expect similar level of static disorder for Se atoms, which is the other member of the Cu-Se plane. In addition, we observed similarly large (albeit slightly smaller) $\sigma^2$ for Ag regardless of doping concentration. Since it is generally expected that doping could cause structural disorder, these observations support the idea that the disorder is not due to static (i.e. structural) disorder. 

There exists a mechanism that could explain the quasi-static origin of the observed thermal disorder of Cu/Ag. To explain the intrinsic low lattice thermal conductivity of BiCuSeO, Vaqueiro \textit{et al.} proposed a ``rattling'' mode mechanism~\cite{Vaqueiro2015}. That is, a localized, independent vibrational ``rattling” mode at Cu-site could scatter phonons, lowering thermal conductivity. The implication is that Cu/Ag is subject to a “flat” potential, making it easy for it to be displaced from the equilibrium position. We could imagine the displacement is incoherent and slow, showing up as quasi-static disorder in our thermal parameter measurements. Of course, as pointed out by previous studies for clathrates~\cite{Koza2008}, careful examination of phonon dispersion is required to confirm presence of a ``rattling'' mode; however, our result indicates such a localized mode (or local disorder) as the origin of low thermal conductivity as discussed next.

The origins of intrinsic low lattice thermal conductivity in BiCuSeO have been debated to date. One of the proposed explanation suggested that the heavy Bi atom is responsible~\cite{Saha2015,SKSaha2016}, with the higher atomic mass leading to lower phonon frequencies and slower sound velocities, reducing heat transfer rate. Another prediction was that flat phonon branches associated with Cu atoms effectively block heat transfer~\cite{DDFan2017,Vaqueiro2015}. The flat phonon branches have smaller group velocities and smaller phonon lifetimes, thus restricting thermal transport~\cite{Li2015}. Since the flat phonon branches are limited to Cu, it suggests that a localized mode on Cu is responsible for these flat phonon branches. This is the mechanism proposed by Vaqueiro \textit{et al.}, a low energy, independent vibrational ``rattling" mode on Cu atom, and our experimental results seem to support this scenario. 

Another important finding that backs up the above picture is our observation of the lack of lattice expansion with Ag doping. The introduction of Ag dopant in Cu-site does not significantly increase the lattice constants, as seen in the case of BiCu$_{1-x}$Ag$_x$SeO in Table~\ref{tab:table1}. Since Ag$^{1+}$ ion (129 pm) is much larger in size than Cu$^{1+}$ ion (91 pm)~\cite{Shannon1976}, this suggests that the void around Cu-site is large, and the additional empty space accommodates Ag dopant without requiring the lattice to expand. This is in contrast to Bi-site, where the replacement of Bi$^{3+}$ (117 pm) with Pb$^{2+}$ (133 pm)~\cite{Shannon1976} leads to a significant increase in lattice constants, as seen for Bi$_{1-x}$Pb$_x$Cu$_{1-x}$Ag$_x$SeO samples in Table~\ref{tab:table1}. Presence of a large void also supports the idea of a ``rattling" mode at Cu-site in BiCuSeO. 

The introduction of Ag leads to a further decrease of lattice thermal conductivity, in both single~\cite{Tan2014,LiuAg2015,Farooq2017} and dual-doped samples~\cite{Li2019}. Before discussing the possible origin of additional suppression of lattice thermal conductivity in Ag doped samples, we would like to comment on the dopant site issue. 

The dopant site for Pb is generally agreed upon, it is expected to replace Bi in the BiCuSeO framework, as shown in Figure~\ref{fig:fig1_structure}. This has been confirmed via XANES measurements in a previous study~\cite{Liu2016}. The dopant site for Ag, however, has been a matter of debate. Single doping studies of Ag in BiCuSeO have found Ag in both Cu and Bi sites~\cite{Tan2014,Farooq2017,Sun2017,LiuAg2015}. A previous (Pb,Ag)$-$dual doping study suggested that some of the Ag dopants did not enter the lattice and produced inter-layer impurities~\cite{Li2019}. Our Ag K-edge EXAFS results show no evidence of any alternate dopant locations, except Cu-site. Hence, within measurement uncertainty, the results point to Ag replacing Cu in BiCuSeO. 

Given that Ag doping does not increase static disorder parameter in our doped samples, we can conclude that the dopant is more or less uniformly (and randomly) replacing Cu atoms. The mass defect therefore given by Ag replacing Cu, coupled with the retained disorder of Ag dopant, could be the main reason for the reduced thermal conductivity. 

It is also worth noting that, while large $\sigma_j^2$ values are observed along the Ag-(Cu/Ag) and Cu-(Cu/Ag) atom-pair in EXAFS results, the values of $\sigma_j^2$ along Ag-Se and Cu-Se are found to be considerably smaller. This result is somewhat unexpected. One possibility is that, within the Cu-Se plane shown in Figure~\ref{fig:fig1_structure}, the Se `sub-layer' is relatively static, while the Cu-site atoms vibrate in-plane with large amplitude. The Se-Cu-Se bond angle changes in the CuSe$_4$ tetrahedron during these vibrations, allowing for smaller changes in Cu-Se bond length (smaller $\sigma_j^2$ along Cu-Se) and smaller vibrations of Se (smaller $U_{iso}$ value of Se). Nevertheless, the role of Se in the Cu-Se plane of BiCuSeO is unclear and should be clarified by further study. 

\section{\label{sec:conclude}Conclusion}
This work presents X-ray diffraction (XRD) and EXAFS measurements on undoped, Ag doped and (Pb,Ag)$-$dual doped BiCuSeO. The main observation is the unusually large isotropic mean-square displacement of Cu in undoped BiCuSeO in XRD results, and the large atom-pair mean-square displacement values along Cu-(Cu/Ag) and Ag-(Cu/Ag) in all samples in EXAFS. These observations indicate the presence of significant intrinsic disorder in Cu-site of BiCuSeO, which is retained by Ag dopant on replacing Cu. No increase of lattice constants is observed on Ag doping at room temperature, suggesting the presence of a large void around Cu-site in BiCuSeO. The temperature independence and presence of large void supports the ``rattling'' mode scenario, localized on Cu-site. This may explain the intrinsic low lattice thermal conductivity of BiCuSeO. The mass defect on introduction of Ag, coupled with the retention of ``rattling'' behaviour, can also explain the further lowering of lattice thermal conductivity of BiCuSeO on Ag doping. The combination of XRD and EXAFS, used to comprehensively understand the structure and dynamics of undoped and doped BiCuSeO, is shown to be a powerful technique which can be applied to the study of other materials with interesting crystal structure and doping properties. 

\begin{acknowledgments}
The work at the University of Toronto was supported by the Natural Sciences and Engineering Research Council (NSERC) of Canada [RGPIN-2019-06449 and RTI-2019-00809], Canada Foundation for Innovation, and the Ontario Research Fund. Research performed at the Canadian Light Source is supported by the Canada Foundation for Innovation, NSERC, the University of Saskatchewan, the Government of Saskatchewan, Western Economic Diversification Canada, the National Research Council Canada, and the Canadian Institute of Health Research. S.J. and D.S.S. acknowledges the receipt of support from the CLS Graduate and Post-Doctoral Student Travel Support Program. B.W.L. acknowledges support from the University of Toronto Faculty of Arts and Sciences Postdoctoral Fellowship and NSERC [funding reference number PDF-546035-2020].
\end{acknowledgments}

\nocite{Ravel}

\providecommand{\noopsort}[1]{}\providecommand{\singleletter}[1]{#1}%

\end{document}


\preprint{}

\title{Investigation of Cu-site disorder in undoped and doped BiCuSeO\\Supplemental Material}

\author{Sheetal Jain}
\author{Christopher J. S. Heath}
\author{Dixshant Shree Shreemal}
\author{Blair W. Lebert}
\affiliation{Department of Physics, University of Toronto, Toronto, Ontario, M5S 1A7, Canada}
\author{Ning Chen}
\author{Weifeng Chen}
\affiliation{Canadian Light Source (CLS), Saskatoon, Saskatchewan, S7N 2V3, Canada}
\author{Young-June Kim}
\affiliation{Department of Physics, University of Toronto, Toronto, Ontario, M5S 1A7, Canada}

\date{\today}

\maketitle

\clearpage

\section{Phase purity analysis using XRD}

\begin{figure*}[htbp]
\centering
\begin{subfigure}[b]{0.48\textwidth}
\includegraphics[width=\linewidth]{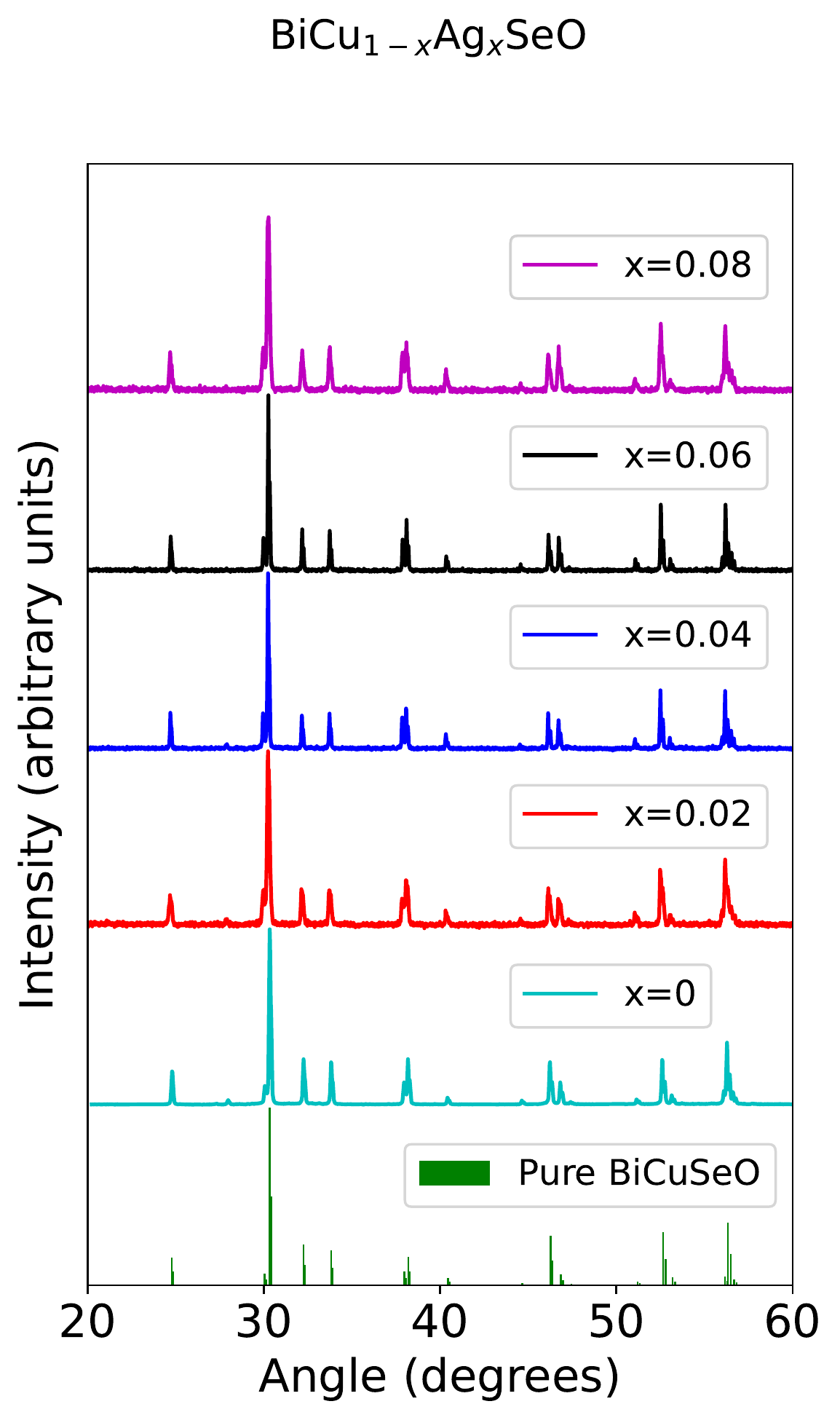}
\caption{\label{fig:XRD1}}
\end{subfigure}
\begin{subfigure}[b]{0.465\textwidth}
\includegraphics[width=\linewidth]{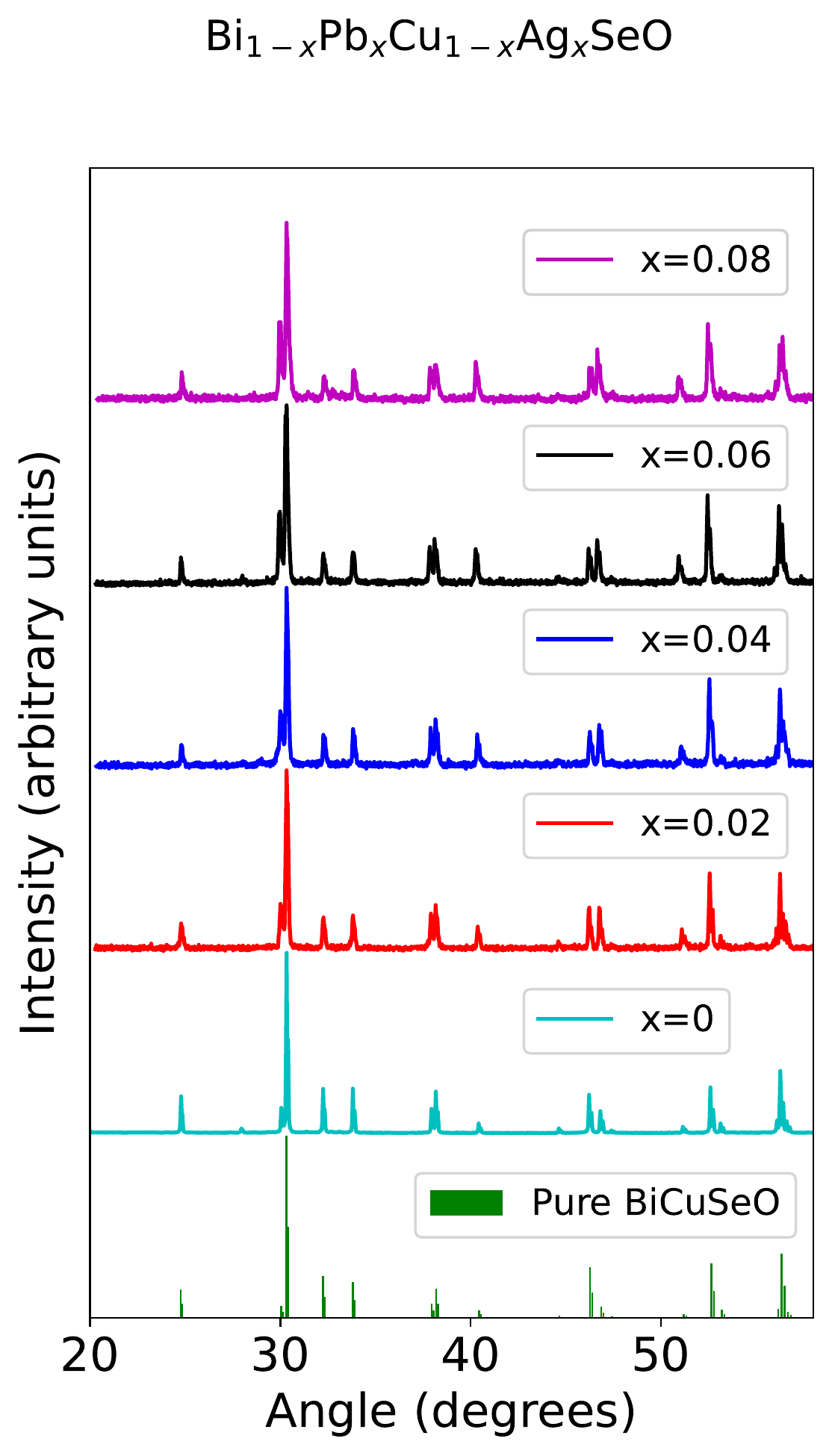}
\caption{\label{fig:XRD2}}
\end{subfigure}
\caption{XRD raw data for \subref{fig:XRD1}~Ag doped and \subref{fig:XRD2}~(Pb,Ag)$-$dual doped BiCuSeO. XRD raw data for undoped BiCuSeO (cyan), as well as a VESTA~\cite{[{Ref. 69, }]Momma2011} simulation of the expected powder XRD pattern for pure BiCuSeO (green barplot), are shown at the bottom of both plots for reference. All XRD peaks for the doped samples are seen to match the peaks obtained for the undoped sample, confirming phase purity. The primary impurity XRD peak of unreacted Bi$_2$O$_3$ (less than 0.5\%) is seen near 27 deg for some samples.}
\end{figure*}

\clearpage

\section{Raw XRD data for undoped BiCuSeO}

\begin{figure}[htbp]
\centering
\includegraphics[width=\textwidth]{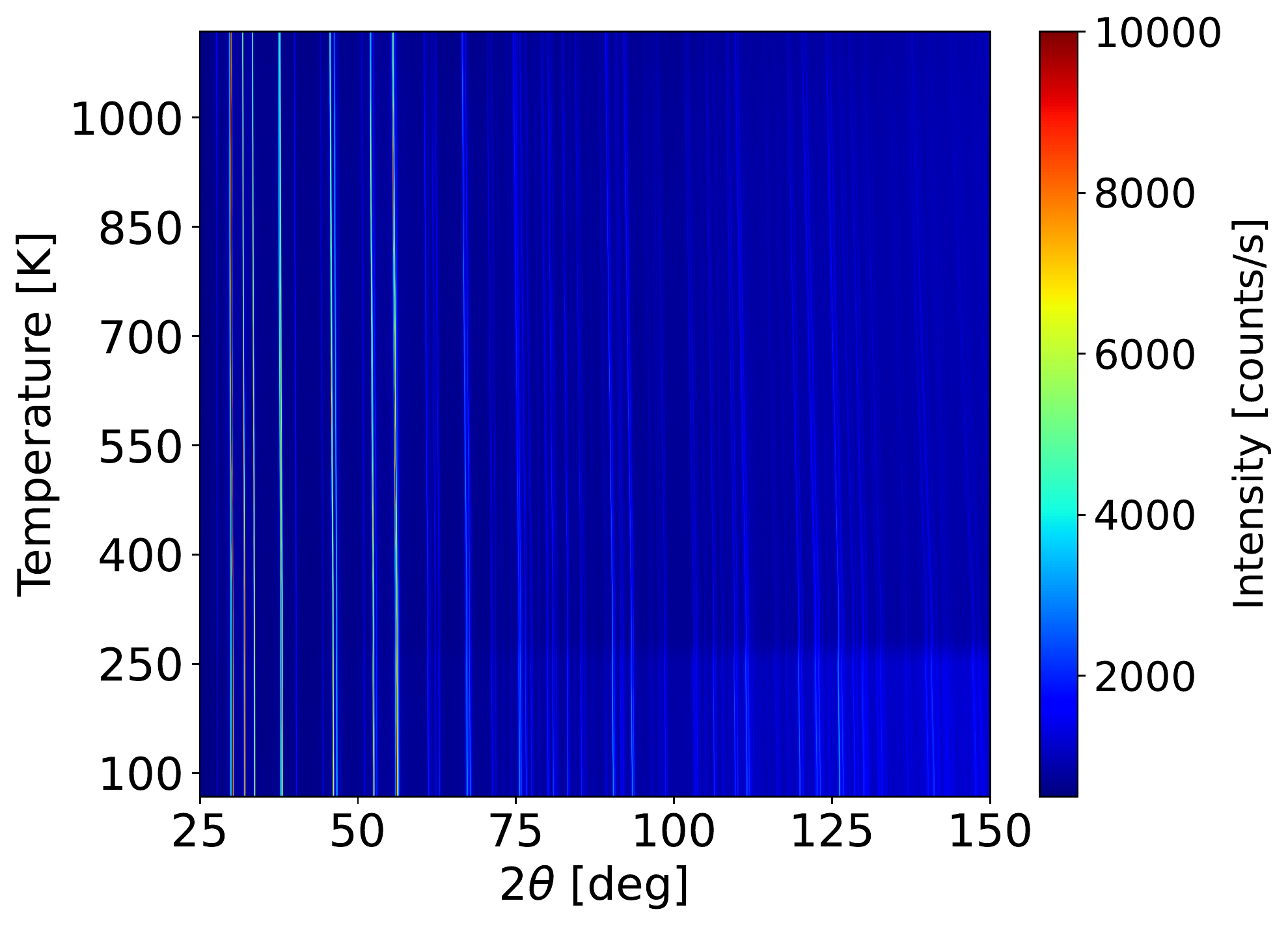}
\caption{\label{fig:XRD}XRD raw data for undoped BiCuSeO, as a function of temperature, for the entire 2$\theta$ range measured. Note the suppression of intensity at high angles due to Debye-Waller factor at high temperatures, seen in the top right of the figure.}
\end{figure}

\clearpage

\section{EXAFS Data and fits}

\begin{figure*}[htbp]
\centering
\begin{subfigure}[b]{0.45\textwidth}
\includegraphics[width=\linewidth]{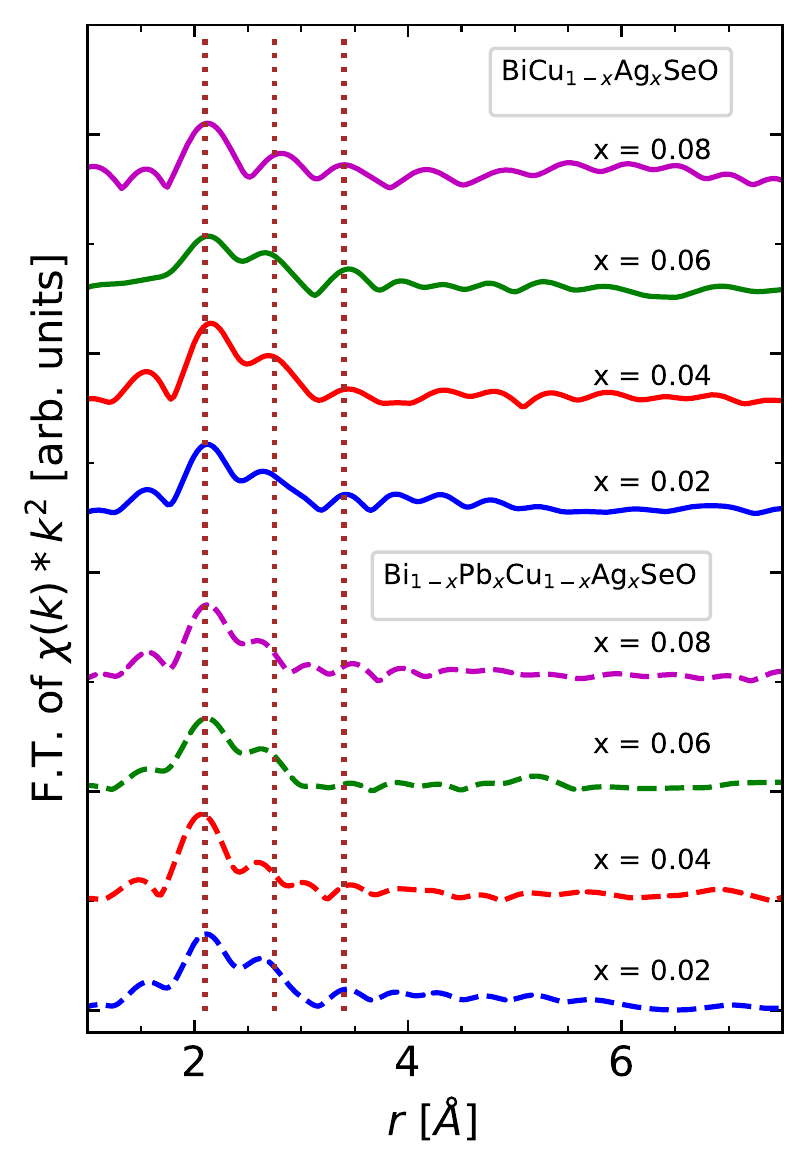}
\caption{\label{fig:Ag1}}
\end{subfigure}
\begin{subfigure}[b]{0.45\textwidth}
\includegraphics[width=\linewidth]{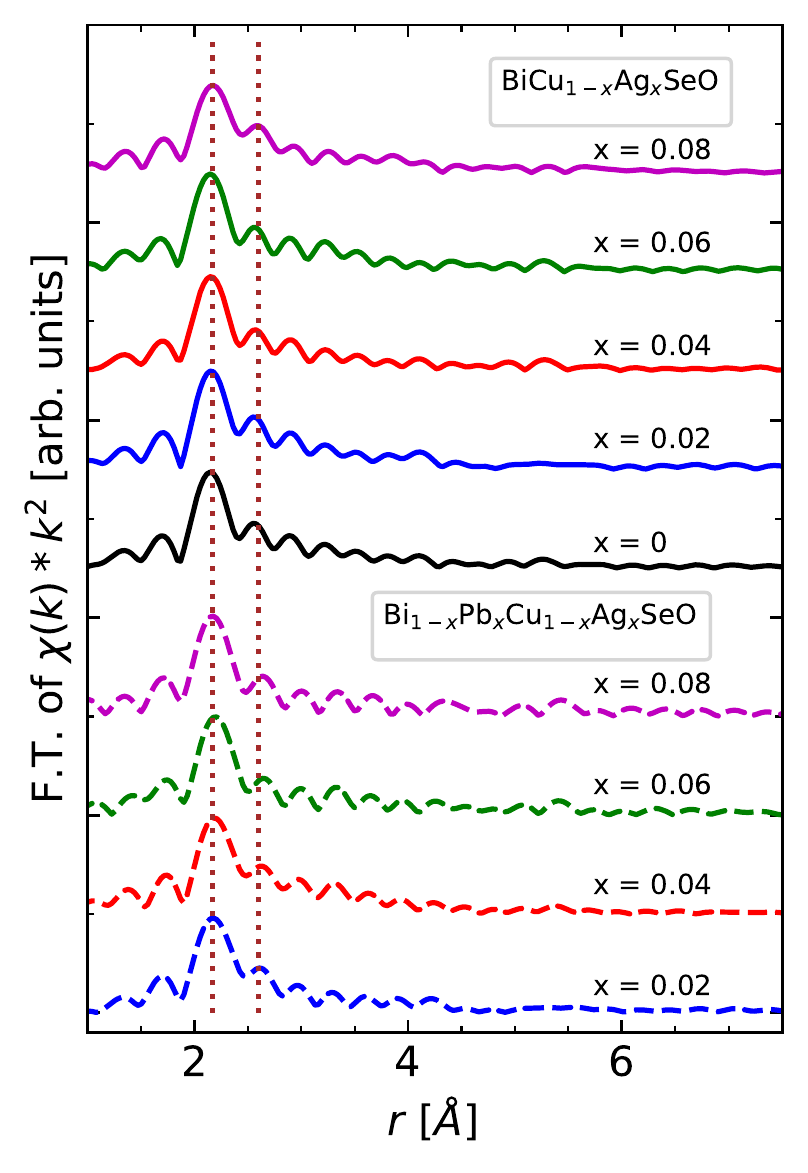}
\caption{\label{fig:Ag2}}
\end{subfigure}
\caption{\subref{fig:Ag1} Ag K-edge EXAFS data in $r$-space for Ag doped (solid lines) and (Pb,Ag)$-$dual doped BiCuSeO (dashed lines). The vertical brown dotted lines represent the first (Ag-Se), second (Ag-(Cu/Ag)) and third (Ag-(Bi/Pb)) shell locations in $r$-space, corresponding to Ag replacing Cu in the BiCuSeO crystal structure framework. The Fourier transform (F.T.) ranges are 3$-$11 \textup{\AA}$^{-1}$. No significant doping dependence is observed in the data. \subref{fig:Ag2}~Cu K-edge EXAFS data in $r$-space for undoped, Ag doped (solid lines) and (Pb,Ag)$-$dual doped BiCuSeO (dashed lines). The vertical brown dotted lines represent the first (Cu-Se) and second (Cu-(Cu/Ag)) shell locations in $r$-space. The Fourier transform (F.T.) ranges are 3$-$14 \textup{\AA}$^{-1}$. No significant doping dependence is observed in this data as well.}
\end{figure*}

\clearpage

\begin{figure*}[htbp]
\centering
\begin{subfigure}[b]{0.45\textwidth}
\includegraphics[width=\linewidth]{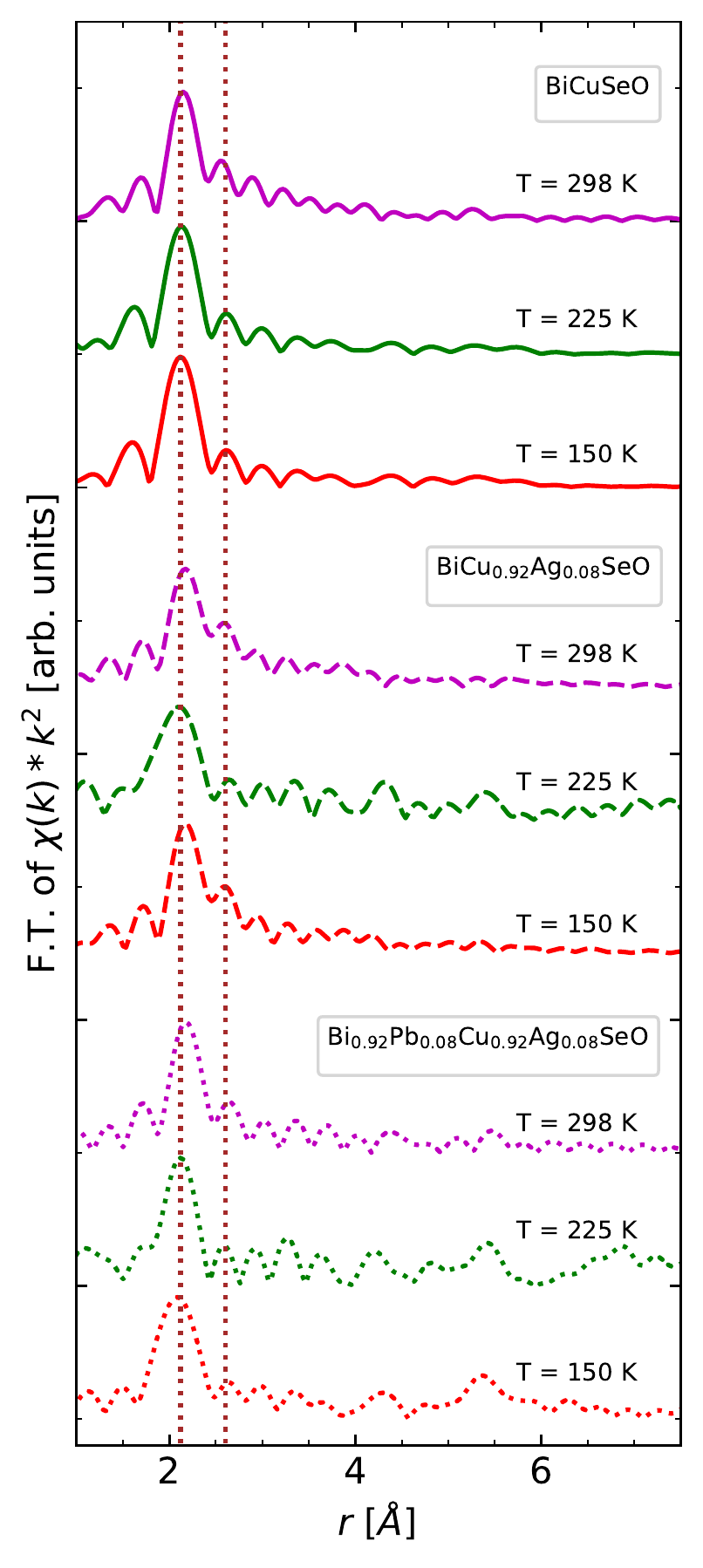}
\caption{\label{fig:Cu1}}
\end{subfigure}
\begin{subfigure}[b]{0.45\textwidth}
\includegraphics[width=\linewidth]{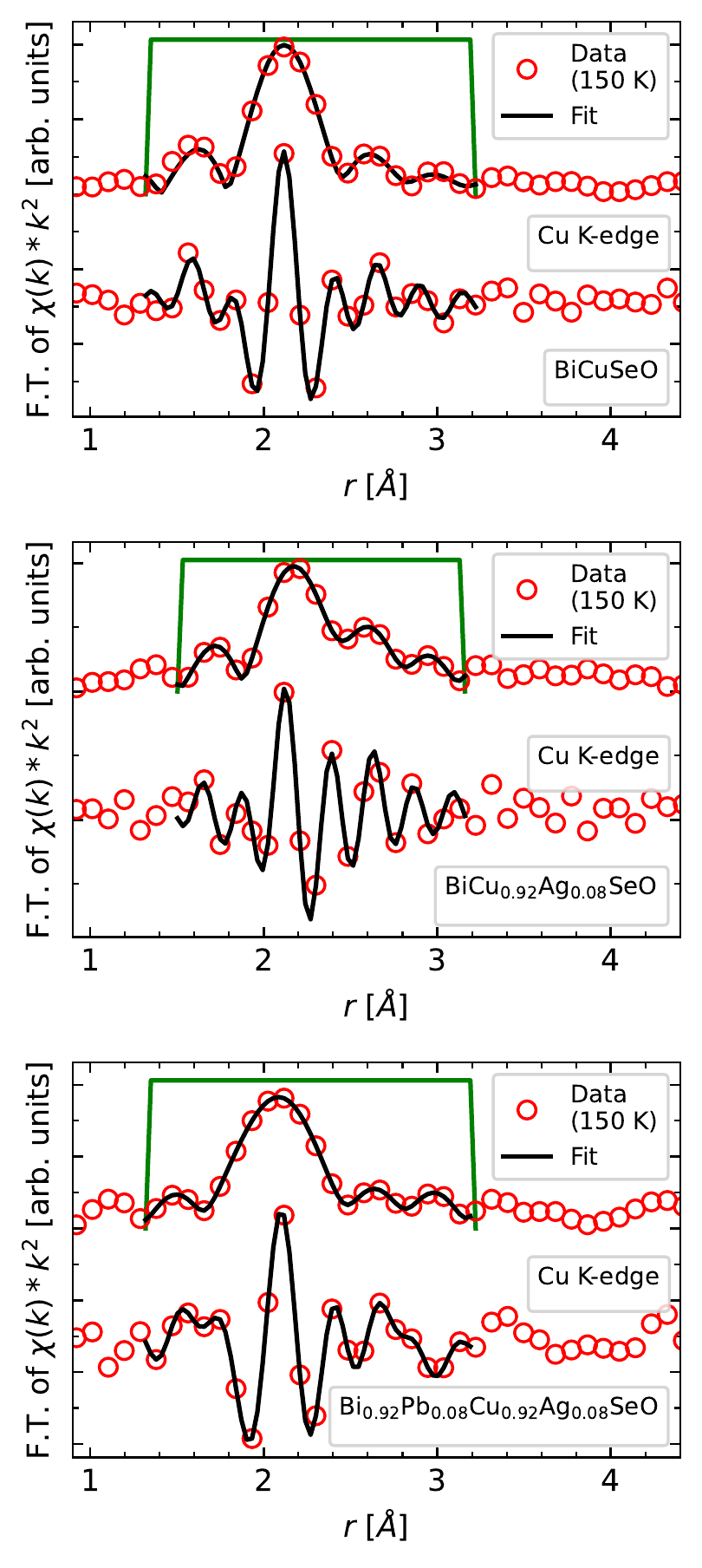}
\caption{\label{fig:Cu2}}
\end{subfigure}
\caption{\subref{fig:Cu1} Cu K-edge EXAFS data in $r$-space for BiCuSeO as solid lines, BiCu$_{0.92}$Ag$_{0.08}$SeO as dashed lines, and Bi$_{0.92}$Pb$_{0.08}$Cu$_{0.92}$Ag$_{0.08}$SeO as dotted lines, at different temperatures. The vertical brown dotted lines represent the first (Cu-Se) and second (Cu-(Cu/Ag)) shell locations in $r$-space. The Fourier transform (F.T.) ranges are 3$-$14 \textup{\AA}$^{-1}$. Note that due to higher noise levels, due to condensation effects at low temperature, EXAFS data at 80 K was not usable. \subref{fig:Cu2} The fit results for BiCuSeO, BiCu$_{0.92}$Ag$_{0.08}$SeO and Bi$_{0.92}$Pb$_{0.08}$Cu$_{0.92}$Ag$_{0.08}$SeO at 150 K. The fit ranges are shown via the window function in solid green lines.}
\end{figure*}

\clearpage

\begin{table*}[htbp]
\caption{\label{tab:table2}Change in atom-pair distance ($\Delta r_j$) along atom-pairs of Ag-Se ($r_1 = 2.442$ \textup{\AA}), Ag-(Cu/Ag) ($r_1 = 2.887$ \textup{\AA}) and Ag-(Bi/Pb) ($r_1 = 3.794 $\textup{\AA}) for Ag doped BiCuSeO and (Pb,Ag)$-$dual doped BiCuSeO samples at different temperatures, obtained from fitting of Ag K-edge EXAFS data.}
\begin{ruledtabular}
\begin{tabular}{ccccccc}
Sample&\multicolumn{3}{c}{BiCu$_{0.92}$Ag$_{0.08}$SeO}&\multicolumn{3}{c}{Bi$_{0.92}$Pb$_{0.08}$Cu$_{0.92}$Ag$_{0.08}$SeO}\\
Temperature [K] & $\Delta r_1$ [\textup{\AA}] & $\Delta r_2$ [\textup{\AA}] & $\Delta r_3$ [\textup{\AA}] & $\Delta r_1$ [\textup{\AA}] & $\Delta r_2$ [\textup{\AA}] & $\Delta r_3$ [\textup{\AA}]\\ \hline
80 & 0.007(7) & 0.007(6) & 0.004(1) & 0.006(6) & 0.006(6) & 0.004(1)\\
150 & 0.008(8) & 0.006(5) & 0.003(1) & 0.008(7) & 0.007(6) & 0.002(1)\\
225 & 0.006(5) & 0.007(7) & 0.005(2) & 0.007(7) & 0.007(6) & 0.003(2)\\
298 & 0.008(7) & 0.006(5) & 0.002(1) & 0.006(5) & 0.006(6) & 0.005(2)\\
\end{tabular}
\end{ruledtabular}
\end{table*}

\section{Goodness of fit for EXAFS}

Generally, EXAFS fits rarely yield reduced-$\chi^2$ close to 1, despite data being consistent to the fitting model~\cite{[{Ref. 87, }]Ravel}. The reason is that the measurement uncertainty is underestimated, yielding large reduced-$\chi^2$ values. The assumption that measurement uncertainty is mostly dominated by statistical variation doesn't hold good in case of the large fluxes of synchrotron radiation. Instrumental sources of uncertainty and error in theoretical estimates of scattering factors are more significant, but estimating and incorporating them into reduced-$\chi^2$ is difficult. 

However, in the case of fluorescence measurements on very dilute samples, statistical variation is indeed the largest source of measurement uncertainty, as is the case for our Ag K-edge EXAFS data. Thus, our fits for that data give reduced-$\chi^2$ close to 1, as is expected for good fits!

\bibliography{suppbib}